\documentclass[10pt, conference]{IEEEtran}

\IEEEoverridecommandlockouts % needed to identify funding in the first footnote
\usepackage{amsthm}
\newtheorem{definition}{Definition}
\usepackage{algorithm}
\usepackage{cite}
\usepackage{amsmath,amssymb,amsfonts}
\usepackage{algorithmic}
\usepackage{graphicx}
\usepackage{textcomp}
\usepackage{xcolor}
\usepackage{caption}
\usepackage{subcaption}
\usepackage{array}
\usepackage{cleveref}
\usepackage [english]{babel}
\usepackage [autostyle, english = american]{csquotes}
\usepackage{tikz}

\MakeOuterQuote{"}
\newcommand\copyrighttext{%
  \footnotesize \textcopyright 2023 IEEE. Personal use of this material is permitted.
  Permission from IEEE must be obtained for all other uses, in any current or future
  media, including reprinting/republishing this material for advertising or promotional
  purposes, creating new collective works, for resale or redistribution to servers or
  lists, or reuse of any copyrighted component of this work in other works.}
\newcommand\copyrightnotice{%
\begin{tikzpicture}[remember picture,overlay]
\node[anchor=south,yshift=10pt] at (current page.south) {\fbox{\parbox{\dimexpr\textwidth-\fboxsep-\fboxrule\relax}{\copyrighttext}}};
\end{tikzpicture}%
}
\begin{document}

\title
{
    A Real-Time Limit Order Book as a Market Mechanism for Transactive Energy Systems

    \thanks
    {
        This work was funded by the US Department of Energy's Office of Energy Efficiency and Renewable Energy Connected Communities program. SLAC National Accelerator Laboratory is operated by Stanford University for the US Department of Energy under Contract DE-AC02-76SF00515.
    }
}

\author{
    
    \IEEEauthorblockN{Akshay Sreekumar}

    \IEEEauthorblockA
    {
        \textit{Electrical and Computer Engineering} 
    \\
        \textit{University of California, Los Angeles}
    \\
        akshay81@g.ucla.edu
    }

    \and

    \IEEEauthorblockN{Adhithyan Sakthivelu}

    \IEEEauthorblockA
    {
        \textit{Civil and Environmental Engineering} 
    \\
        \textit{Stanford University}
    \\
        adhiths@stanford.edu
    }

    \and

    \IEEEauthorblockN{Rimvydas Baltaduonis}

    \IEEEauthorblockA
    {
        \textit{Department of Economics} 
    \\
        \textit{Gettysburg College} 
    \\
        rbaltadu@gettysburg.edu
    }
    
    \and

    \IEEEauthorblockN{Lynne Kiesling}

    \IEEEauthorblockA
    {
        \textit{Institute for Regulatory Law and Economics} 
    \\
        \textit{University of Colorado Denver} 
    \\
        lynne@knowledgeproblem.com
    }

    \and
    
    \IEEEauthorblockN{Seth Hoedl}

    \IEEEauthorblockA
    {
        \textit{Post Road Foundation} 
    \\
        shoedl@postroadfoundation.org
    }

    \and

    \IEEEauthorblockN{David P. Chassin, \textit{Senior Member, IEEE}}

    \IEEEauthorblockA
    {
        \textit{Grid Integration Systems and Mobility} 
    \\
        \textit{SLAC National Accelerator Laboratory}
    \\
        dchassin@slac.stanford.edu
    }
}

\maketitle
\copyrightnotice

\begin{abstract}

    This paper presents a limit order book (LOB) market mechanism design for transactive energy systems. The proposed design is planned for deployment in New Hampshire and Maine under a US Department of Energy Connected Communities project. The new LOB mechanism is intended to replace or work in conjunction with the conventional transactive energy double auction mechanisms designed for retail real-time electricity price discovery, and will facilitate significant scaling of transactive energy systems. The paper provides LOB market rules, clearing algorithm, and illustrative examples and discusses clearing algorithm performance and reliability. The proposed LOB design includes support for discovering prices arising from wholesale electricity markets, distribution system asset constraints, distributed energy resource constraints, and consumer willingness to consume or produce at a reservation price.  

\end{abstract}


\section{Introduction} \label{sec:introduction}

    The increasing penetration of distributed energy resources (DERs) like wind, solar, and electric vehicles introduces stochasticity into supply and demand that can make centralized optimal dispatch difficult at large scale. Efficient coordination of these decentralized grid assets is necessary to increase the flexibility of electric power systems when a high fraction of DERs are present. By employing digital technologies and simultaneously marshaling DERs that produce and consume energy, grid operators can harness flexibility to increase the rate of energy infrastructure decarbonization.
    
        \begin{figure}[!b]

        \centerline{\includegraphics[width=.95\columnwidth]{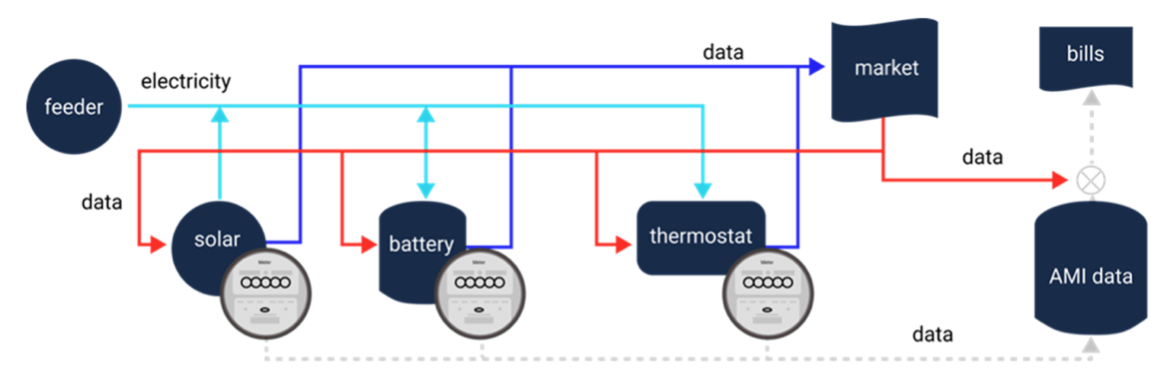}}

        \caption{An overview of a transactive energy system. Devices buy and sell electricity by reacting to price signals discovered through a market mechanism.} 
        \label{fig:TESSDiagram}

    \end{figure}

    Transactive energy systems \cite{ambrosio2016} were originally proposed to coordinate devices on the grid. Devices respond to real-time prices and the prices can either be synthesized by DSOs in advance based on past cost-based responses and forecasts (Type 1 systems) or discovered using retail market mechanisms that rely on device asks and offers (Type 2 systems) based on consumer preferences and expectations. Real-time energy pricing for consumers and producers can coordinate DERs on the grid while offering benefits to all participants in the market, including utilities. Transactive energy supports peer-to-peer trading whereby all market participants can benefit, while providing the utility superior coordination that can manage capacity constraints and energy reserves on the distribution side.

    In past US DOE demonstrations \cite{hammerstrom2008,widergren2013}, transactive energy systems have relied on a time-delimited double auction (DA) market mechanism that clears periodically to discover retail real-time prices. The DA is an information-rich market design, but the market clearing interval is strongly related to the time constant of the devices' response to price signals, which limits the types of devices that can participate to only those with time constants more than 5-10 times the clearing interval. The period's dependence on the number and nature of participating devices limits system responsiveness and the total number of devices. Devices with physical operating timescales that differ significantly from the market clearing period are forced to operate on the auction's timescale, which may not match the bandwidth of the device's maximum response. 
    
    To address these issues, here we consider a real-time limit order book (LOB) mechanism instead of the time-delimited double auction for Type 2 transactive energy systems \cite{weber2011}. A LOB enables the market and the devices to respond to real-time prices by clearing the market as a series of asynchronous double auctions. Device can submit orders at a timescale consistent with the physical operation of the device, thus allowing for efficient markets. Although there have been studies of using an orderbook to coordinate DERs \cite{Qin2018}, this paper introduces an orderbook market mechanism to do so in a transactive energy system setting.

    In Section~\ref{sec:orderbook} we present the market design, participation model, device model, order and dispatch structure, and the order priority rules. In Section~\ref{sec:algorithm} we present the order-book matching algorithm and the consequences of various incentives. In Section \ref{sec:settlement} we discuss the settlement structure of how devices are paid for fulfilled orders. In Section \ref{sec:example} we consider an example of the matching algorithm and settlement, and in Section \ref{sec:conclusion} we offer closing remarks and future directions for research.

\section{The Limit Orderbook} \label{sec:orderbook}

\begin{figure}[!t]

        \centerline{\includegraphics[width=0.75\columnwidth]{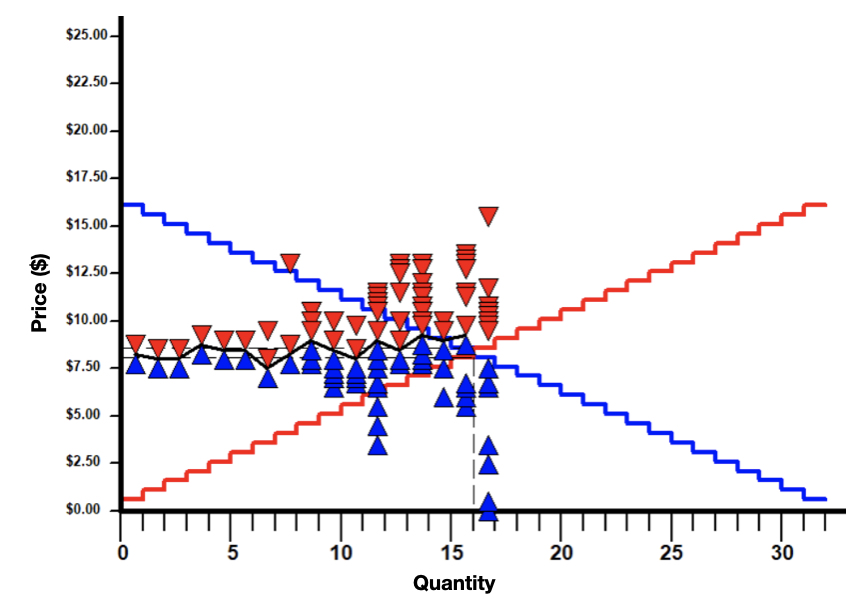}}
        \caption{A sequential double auction model showing convergence to equilibrium quantity and price. Blue denotes buyer bids and red denotes seller asks.} 
        \label{fig:DAmodel}

    \end{figure}

    Both the transactive DA and the transactive LOB are double auctions. A transactive LOB is distinguished from a transactive DA by the following features. First, a DA is simultaneous and cleared at a regular interval, whereas a LOB is sequential (Fig. \ref{fig:DAmodel}) and cleared immediately when a new order is matched. Second, all unfilled orders in a cleared DA are deleted and devices must resubmit their orders for each auction, whereas uncleared orders remain in the LOB for as long as the devices need.  Third, most DA implementations do not allow for partial order clearing, resulting in dispatch errors, whereas the LOB allows partial order filling for flexible devices.  Finally, a DA does not need to recognize time-priority in order filling, whereas a LOB must use time-priority to incentivize early and complete order submission. 

    A key feature of the LOB is that it enables hierarchical market structures that can be aligned with the constraints of a radial distribution network without requiring inter-auction clearing or intertemporal coordination. When a LOB receives an order that is cleared with a resource coming from the market above, it can immediately satisfy the order using a market order. This facilitates fast and easy coordination among multiple adjacent markets and allows for significantly larger deployments of transactive energy systems than possible with a DA structure. However, the focus of this preliminary work is to study the LOB operation in a localized region for a single feeder, where the distribution constraints and losses do not play a major role. 
    
    \subsection{Basic Orderbook Design Rules}

 Buyers and sellers both submit \textit{orders}, with a buy order hereby known as a \textit{bid} and a sell order known as an \textit{ask}. A LOB is defined by the tuple of ordered lists ($\mathcal{B}, \mathcal{S}$), where $\mathcal{B}$ denotes the bids and $\mathcal{S}$ denotes the asks. These bids and asks are sorted by \textit{price-time priority}, such that the oldest, most competitively priced orders on each side will be at the top of the list. The order book, through the use of a matching algorithm (discussed in Section III), can match individual buyers and sellers when their bids and asks are priced favorably (\Cref{fig:LOBPlotExample}). The result of the matching operation will produce a \textit{dispatch} that specifies transactions that need to be carried out between cleared devices. The LOB also allows two different distinct types of orders: \textit{limit orders} and \textit{market orders}. Limit orders specify a reservation price above/below which the order will not match with a corresponding order. Limit orders can have a specified expiration time such that they exist in the LOB until they expire or are otherwise canceled by the user. A market order is intended to execute immediately at the current market price, which is/are the standing order/s on the other side that has the highest price-time priority. 
 % Adhi: Think the above statement explains the concept of market order and the following example is not needed? 
%  A market order can be thought of as a limit order with a limit price of $\infty$ for buy orders, and $0$ for sell orders—the order such that they immediately match with the best available order. 
 We further detail the specific structure and characteristics of these order types in Subsection~\ref{sec:orders}.

    \subsection{Market Participants}

    We consider a set of \textit{devices} that participate in the LOB market by placing buy and sell orders. Below, we enumerate the set of devices and give a brief summary how they would participate based on their features and the motivations of their owners/operators.

    \begin{itemize}
    
        \item \textbf{HVAC}: Minimize costs while maintaining building indoor air temperature near a desired temperature setpoint.
        
        \item \textbf{Water Heater}: Minimize costs while maintaining the hot water near a desired temperature setpoint.
        
        \item \textbf{EV Chargers (V0G, V1G)}: Minimize costs while ensuring vehicle is charged to the desired state of charge (SOC) by the desired time.
        
        \item \textbf{EV Charger (V2G)}: Maximize profit while ensuring vehicle is charged to the desired state of charge (SOC) by the desired time.
        
        \item \textbf{PV}: Maximize revenue by selling excess energy.
        
        \item \textbf{Battery}: Maximize profit while maintaining a desired level of charge for occupants' reliability/resilience needs.
        
        \item \textbf{Feeder/Utility}: Bids the wholesale price of electricity in bid orders or the locational marginal price in ask orders to act as a pseudo-market maker providing liquidity in the market
        
    % Adhi: Feeder in bid orders? Are we talking about the hierarchical structure? 
    % ^ This is the same content/text that was previously in a separate section, just formatting wise has moved here
    \end{itemize}

    \subsection{Device Flexibility}

    \textit{Flexibility} is the notion that some devices can accept partial fulfillment of their order quantity based on the devices' physical constraints. A power \textit{inflexible} device has to produce or consume the exact quantity of power in it's order. On the other hand, a power \textit{flexible} device accepts a match for a quantity of power less than its order. 
    % The notion of flexibility only exists in the quantity dimension, not in time. All orders are assumed to be time flexible, meaning that they will match with an order even if the duration of the orders are not matched. 
    
    % Adhi: should we even introduce the concept of time flexibility?

    For example, the majority of HVAC systems have a nominal operating power. Most HVAC units cannot operate at lower powers\cite{AFRAM2014507}, and thus these units can only place inflexible orders in the market. The list of inflexible devices being considered is: Thermostatic loads (HVAC/hot water), PVs (without curtailment), and V0G chargers. The list of flexible devices is: Batteries, PVs (with curtailment), and V1G/V2G chargers.

    % In Table \ref{device_flexibility_table} we enumerate a set of both power flexible and power inflexible devices currently being considered. 
    
    % The flexibility of an order has significant consequences for the LOB matching algorithm, which will be discussed in greater detail in Section~\ref{sec:algorithm}. 

    % \begin{table}[]
        
    %     \caption{Device Flexibility}
    %     \label{device_flexibility_table}
        
    %     \centering
        
    %     \begin{tabular}{|c|c|}
    %     \hline
    %         \multicolumn{1}{|c|}{\textbf{Power Flexible}} 
    %     & 
    %         \multicolumn{1}{c|}{\textbf{Power Inflexible}} 
    %     \\ \hline
    %         Battery 
    %     & 
    %         Thermostatic loads
    %     \\
    %     &
    %         (HVAC/hot water) 
    %     \\ \hline
    %         PV (w/curtailment) 
    %     & 
    %         PV (w/o curtailment) 
    %     \\ \hline
    %         V1G/V2G 
    %     & 
    %         V0G 
    %     \\ \hline
    %     \end{tabular}
    
    % \end{table}

    \subsection{Order Structure} \label{sec:orders}

    Every order requires that the following set of base attributes be associated with it:

    \begin{itemize}

        \item \textbf{Order ID}: Unique identifier for an order.

        \item \textbf{Device ID}: Unique identifier for device.

        \item \textbf{Quantity}: Amount of power to transact. Positive for buy order. Negative for sell order.

        \item \textbf{Timestamp}: Indicates when an order was received.

        \item \textbf{Duration}: The amount of time for which a device is requesting power.

        \item \textbf{Flexibility}: Boolean. Specifies whether a device is willing to transact partial quantity.

    \end{itemize}

    \noindent A limit order will have all the attributes enumerated above, in addition to these following attributes:

    \begin{itemize}
    
        \item \textbf{Limit Price}: On the buy side, the limit specifies the maximum price one is willing to pay for the quantity. On the sell side, the limit specifies the minimum price at which one is willing to sell the quantity.
    
        \item \textbf{Expiration}: Specifies how long a limit order will exist unmatched in the LOB before it is automatically deleted.
    
    \end{itemize}
% TODO: Price-Time Priority
% 
    \subsection{Price-Time Priority}
    The most fundamental criteria for order matching is the notion of \textit{price-time priority} \cite{Preis2011}. Recall that the LOB consists of the sorted collection of buy and sell orders ($\mathcal{B}, \mathcal{S}$). The orders in $\mathcal{B}$ are sorted such that the orders with the highest bid price are given first priority for matching. Similarly, the orders in $\mathcal{S}$ are sorted such that the orders with the lowest ask price are given first priority for matching. Note that this means all market orders are automatically given the highest priority. To distinguish between orders that have the same limit price (or are both market orders), the order that arrives later to the LOB will be given lower priority. 
    \subsection{Dispatch Structure}
    After orders are matched in the LOB, the relevant transaction information is summarized in a \textit{dispatch}. Dispatches detail how devices should transact power with each other to ensure that their orders are properly fulfilled while respecting the LOB rules. A dispatch can be denoted by a list of transactions where each transaction is given by a tuple of the form ($DeviceId_{seller}$, $DeviceId_{buyer}$, $Quantity$, $Price_{clearing}$, $Duration$).

    % Comment: Shouldn't the dispatch also have duration component?
    % DPC: yes

    % The incentive for price-time priority as the basic structure of matching is clear—orders can jockey for position in the LOB by adjusting their marginal value for power, reflected in the price of the order.

\section{Orderbook Matching Algorithm} \label{sec:algorithm}

    % An efficient matching algorithm underlies the operation of any orderbook. 
    The efficient matching of sellers with buyers in real-time drives liquidity in the market and ensures that devices can operate at their optimal response bandwidth. 

    \subsection{Orderbook Equilibrium}
    
    \begin{definition}

        The best (lowest price) sell order in the LOB is denoted as $\mathcal{S}_{0}$. The best (highest price) buy order in the LOB is denoted as $\mathcal{B}_{0}$. The \textbf{spread} of the LOB is $\mathcal{S}_{0} - \mathcal{B}_{0}$.

    \end{definition}

    \begin{definition}\label{def:stable_LOB}
        The LOB is said to be in equilibrium if either of the following conditions are satisfied:
        \begin{itemize}
            \item The spread is positive; or
            \item The spread is nonpositive but the marginal units are inflexible and the market cannot be cleared and dispatched.
        \end{itemize}
    \end{definition}
    Equilibrium signifies LOB stability—no further matching can occur with the present sets ($\mathcal{B}, \mathcal{S}$). The goal of the matching algorithm is to achieve the LOB equilibrium. 

    % \subsection{Orderbook Visualization}

    % %% show reduced orderbook list, show LOB plot (labeled spread, labeled (B', S'), show quantity stacks

    % An example of a LOB ($\mathcal{B}, \mathcal{S}$) given in tabular form is shown in Table~\ref{tab:LOBTableExample}. An example of a graphical representation of a different LOB is given in Figure~\ref{fig:LOBPlotExample}. Note that the LOB in Figure~\ref{fig:LOBPlotExample} is \textit{not} in equilibrium as it has a spread of -\$10.

    % DPC: wouldn't the green area always only have one order in it, either a buy or a sell?
    % Akshay: Yeah, this graphic is not showing what I want to show. I want to show that you should first just consider the subset of orders where an order could match based on price alone (not considering flexibility). It might be clearer to explain this as a simple algorithm, and then still leave a LOB visual but without the shaded area. 
    % \begin{figure}[b]
    %     \centerline{\includegraphics[width=0.7\columnwidth]{LOB_picture_PES_2.png}}
    %     \caption{Graphical Representation of LOB}
    %     \label{fig:LOBPlotExample}
    % \end{figure}
    
    \begin{figure*}[htbp]
     \centering
     \begin{subfigure}[b]{0.24\textwidth}
         \includegraphics[width=\textwidth]{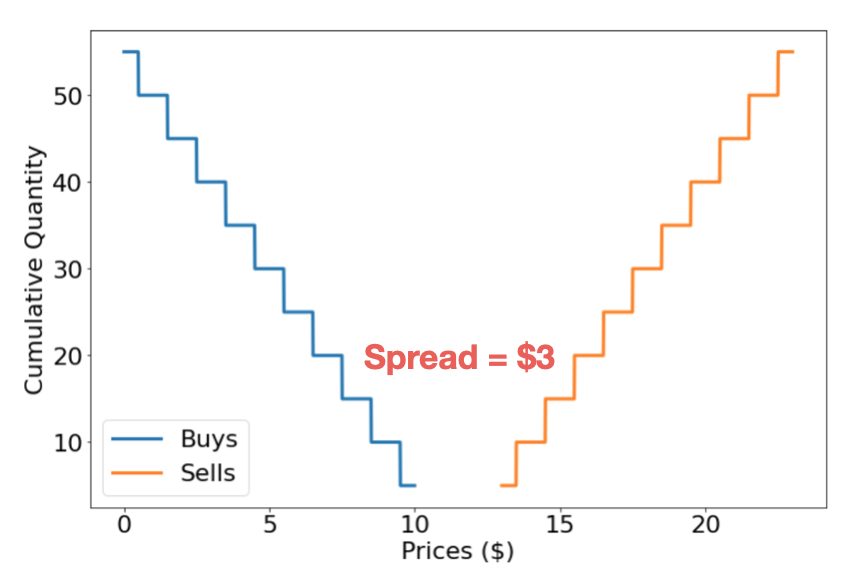}
         \caption{Positive spread equilibrium}
     \end{subfigure}
     \begin{subfigure}[b]{0.24\textwidth}
         \includegraphics[width=\textwidth]{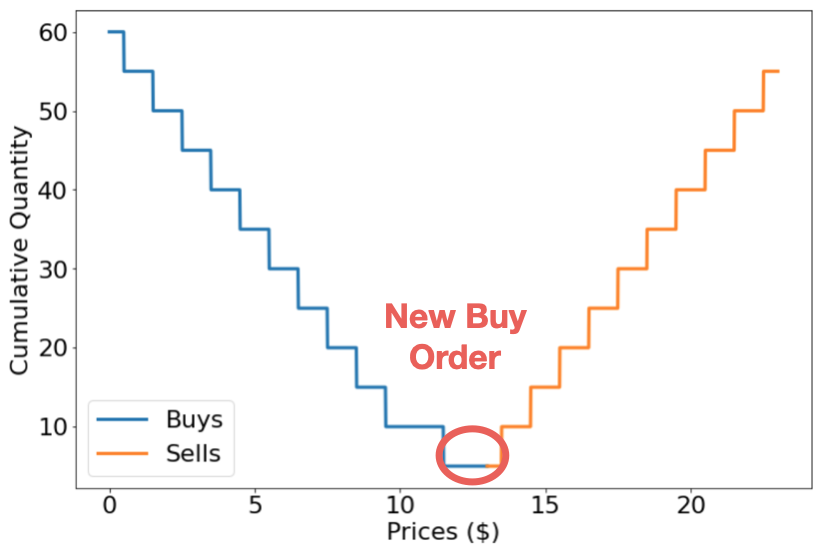}
         \caption{New order dis-equilibrium}
     \end{subfigure}
     \begin{subfigure}[b]{0.24\textwidth}
         \includegraphics[width=\textwidth]{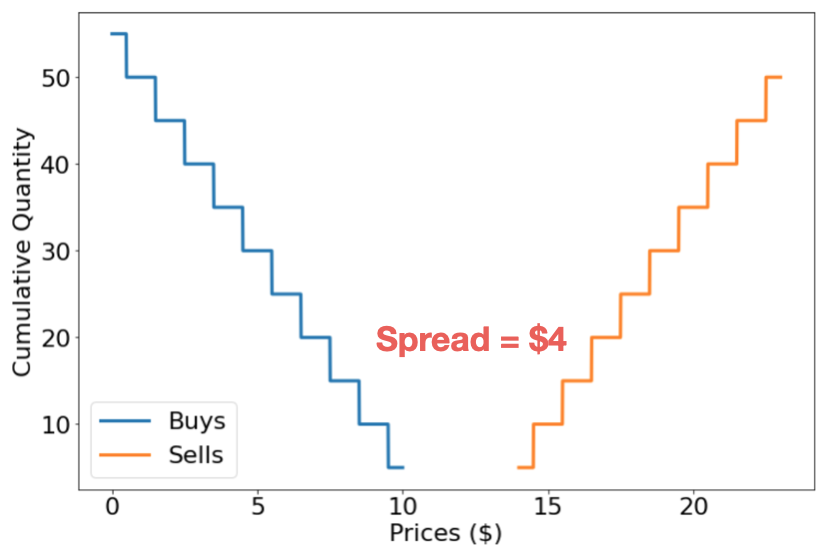}
         \caption{Restored spread equilibrium}
     \end{subfigure}
    \caption{Graphical Representation of LOB}
    \label{fig:LOBPlotExample}
\end{figure*}

    \begin{table*}[]
        \caption{Example Snapshot of LOB (not in equilibrium) in Tabular Form }
        \label{tab:LOBTableExample}
        \centering
        \begin{tabular}{|cccccccc|}
        \hline
            \multicolumn{1}{|c|}{\textbf{Device ID}} 
        & 
            \multicolumn{1}{c|}{\textbf{Order ID}} 
        & 
            \multicolumn{1}{c|}{\textbf{Timestamp}} 
        & 
            \multicolumn{1}{c|}{\textbf{Quantity}} 
        & 
            \multicolumn{1}{c|}{\textbf{Price}} 
        & 
            \multicolumn{1}{c|}{\textbf{isPowerFlexible}} 
        & 
            \multicolumn{1}{c|}{\textbf{Duration}} & \textbf{Expiration} 
        \\ \hline 
            \multicolumn{8}{|c|}{~}
        \\ 
            \multicolumn{8}{|c|}{\textbf{BUY}} 
        \\ \hline
            \multicolumn{1}{|c|}{1} 
        & 
            \multicolumn{1}{c|}{1} 
        & 
            \multicolumn{1}{c|}{2022-01-01 00:00:00} 
        & 
            \multicolumn{1}{c|}{2} 
        & 
            \multicolumn{1}{c|}{4} 
        & 
            \multicolumn{1}{c|}{FALSE} 
        & 
            \multicolumn{1}{c|}{10} 
        & 
            10 
        \\ \hline
            \multicolumn{1}{|c|}{2} 
        & 
            \multicolumn{1}{c|}{2} 
        & 
            \multicolumn{1}{c|}{2022-01-01 00:06:25} 
        & 
            \multicolumn{1}{c|}{2} 
        & 
            \multicolumn{1}{c|}{3} 
        & 
            \multicolumn{1}{c|}{FALSE} 
        & 
            \multicolumn{1}{c|}{10} 
        & 
            10 
        \\ \hline
            \multicolumn{8}{|c|}{~}
        \\
            \multicolumn{8}{|c|}{\textbf{SELL}} 
        \\ \hline
            \multicolumn{1}{|c|}{4} 
        & 
            \multicolumn{1}{c|}{4} 
        & 
            \multicolumn{1}{c|}{2022-01-01 00:12:07} 
        & 
            \multicolumn{1}{c|}{-3} 
        & 
            \multicolumn{1}{c|}{1} 
        & 
            \multicolumn{1}{c|}{FALSE} 
        & 
            \multicolumn{1}{c|}{10} & 10 
        \\ \hline
            \multicolumn{1}{|c|}{3} 
        & 
            \multicolumn{1}{c|}{3} 
        & 
            \multicolumn{1}{c|}{2022-01-01 00:08:18} 
        & 
            \multicolumn{1}{c|}{-2} 
        & 
            \multicolumn{1}{c|}{2.5} 
        & 
            \multicolumn{1}{c|}{TRUE} 
        & 
            \multicolumn{1}{c|}{10} 
        & 
            10 
        \\ \hline
        \end{tabular}
    \end{table*}

    % We are first interested in computing the subsets $\mathcal{B'}\subseteq \mathcal{B}$ and $\mathcal{S'} \subseteq \mathcal{S}$.

    % Note the subset of the LOB ($\mathcal{B'}, \mathcal{S'}$) labeled graphically in Figure~\ref{fig:LOBPlotExample}. We consider the mapping $f: (\mathcal{B}, \mathcal{S}) \mapsto   (\mathcal{B'}, \mathcal{S'})$ to generate this subset. ($\mathcal{B'}, \mathcal{S'}$) denotes the largest possible sets of $\mathcal{B}$ and $\mathcal{S}$ such that all orders contained in the subset are able to match based on price (but not necessarily based on flexibility). This subset will play an important role in the matching algorithm described below. 

    \subsection{Quantity Matching Algorithm}
    
    We first consider the simpler case of \textit{only flexible orders} in the LOB. We assume that there is a LOB ($\mathcal{B}, \mathcal{S}$) in equilibrium with non-empty $\mathcal{B}, \mathcal{S}$. The spread is thus positive. Without loss of generality, consider the arrival of a new ask $S_{i}$ with quantity $Q_{S_{i}}$. There are two possibilities:
    
    \textbf{Case 1}: The spread is still positive even after considering the limit price of $S_{i}$. In this case, $S_{i}$ is added to the LOB based on price-time priority and the LOB remains in equilibrium.
    
    \textbf{Case 2}: The spread is nonpositive after considering the limit price of $S_{i}$, or $S_{i}$ is a market order. In this case, the LOB is no longer in equilibrium and the LOB will fulfill the incoming order. We define $Q_{REMAIN} = Q_{S_{i}}$. We iterate through the buy orders $B_{j} \in \mathcal{B}$ each with quantity $Q_{B_{j}}$ starting with the highest priority buy order and begin fulfilling $Q_{REMAIN}$. If $Q_{REMAIN} \neq 0$ even after iterating through all the potential buy orders, then matching is unsuccessful. If $Q_{REMAIN} = 0$, then matching is successful and the LOB has been returned to equilibrium. As orders are successfully matched with $S_{i}$, the transaction is added to the dispatch. \Cref{fig:LOBPlotExample} illustrates this case.
    
    In the above case, the matching operation is only a function of price. We now consider both flexible and inflexible orders in the LOB, which makes matching a function of both price and flexibility. Flexibility necessitates considering several orders simultaneously in matching because the LOB may be in a nonpositive-spread equilibrium, meaning that the addition of a single order can trigger multiple orders matching on the same side. Thus, we first filter the LOB ($\mathcal{B}, \mathcal{S}$) by price.
    
    \begin{definition}
    \label{PrimeDef}
    
    The price-filtered sets $\mathcal{B'}\subseteq \mathcal{B}$ and $\mathcal{S'}\subseteq \mathcal{S}$ are the \textbf{largest} possible subsets of ($\mathcal{B}, \mathcal{S}$) such that the price of the least competitive ask in $\mathcal{S'}$ is $\leq$ the price of the least competitive bid in $\mathcal{B'}$.
    \end{definition}
    
    In other words, ($\mathcal{B'}, \mathcal{S'}$) denotes the largest possible sets of $\mathcal{B}$ and $\mathcal{S}$ such that all orders contained in the subset are able to match based on price (but not necessarily based on flexibility). We may consider these subsets as "stacks" of orders sorted by priority. Our goal is to match these stacks with each other.

    \begin{definition}
    
        We denote the total quantity demanded represented by all the orders in $\mathcal{B'}$ as $D'_{TOT}$ and the total quantity supplied represented by all the orders in $\mathcal{S'}$ as $S'_{TOT}$.
    
    \end{definition}

    We now consider the following cases for stack clearing given the price-filtered sets ($\mathcal{B'},\mathcal{S'})$.

    \textbf{Case 1}:  $D'_{TOT} = S'_{TOT}$ (\textit{Matched Stacks})

    In this case, every single order in ($\mathcal{B'}, \mathcal{S'}$) will be matched. We iterate through the buy orders beginning with the highest priority buy order. For each buy order, we match it with sell orders (starting with the highest priority sell orders) until the buy order is fulfilled and continue with the next buy order. As both buy and sell orders are fulfilled, the transaction is added to the dispatch to be returned once matching is done.
    
    % Because the total power supplied is equal to the total power demanded, we can add all the orders in the price-filtered set ($\mathcal{B'}, \mathcal{S'}$) directly to dispatch.
    
    \textbf{Case 2}:  $D'_{TOT} \neq S'_{TOT}$ (\textit{Stack Cutting})
    
    In this case, total demand is not equal to total supply. Our goal is to exclude some of the supply or demand, "cutting" the stacks to match quantity, as illustrated in \Cref{fig:StackCutting}. Then, the problem simply reduces to \textbf{Case 1}. To do so, we consider the stack with the greater quantity and check whether the low-priority \textit{marginal} order is flexible. If the marginal order is flexible, then we simply cut enough quantity from the marginal order such that $D'_{TOT} = S'_{TOT}$ and proceed with \textbf{Case 1}. If the marginal order is \textit{not} flexible, then we exclude the entire inflexible order and consider the next order based on priority then recompute $D'_{TOT}$ and $S'_{TOT}$. We repeat this process until \textbf{Case 1} or we run out of orders to exclude. In the latter scenario, we determine that matching is not possible and the LOB lies in equilibrium according to \textbf{Definition} \textbf{\ref{def:stable_LOB}}.
    
    % Adhi: I think we have to specify we are cutting the lower-priority orders, even though marginal specifies the last unit. Making edits to say the same.

    Price-filtering reduces orderbook matching to an asynchronous DA. All devices in the subsets $\mathcal{B'}, \mathcal{S'}$ are potentially in contract with each other, and will all only pay a single price (see Section \ref{sec:settlement} for details). A full example of the matching algorithm is given in Section \ref{sec:example}.

    \subsection{Flexible vs. Inflexible Orders}

    As observed in the matching algorithm, an inflexible order may get skipped over despite having superior price-time priority. Skipping inflexible orders serves two purposes: (i) it incentivizes devices to behave flexibly to promote liquid markets and (ii) attempting to match the inflexible orders with a subset of unmatched orders, while honoring the order's priority, would result in an intractable combinatorial problem.
    
% Adhi: It might not be clear to the readers what backtracking is, should we try to explain it more clearly like "...attempting to match the inflexible orders with a subset of unmatched orders, while honoring the priority of the order would result in an intractable combinatorial problem and ..... " 

    \subsection{Flexible Order Partial Matching}
    \label{sec:partialmatching}

    A flexible order may be fulfilled only partially in both quantity and duration. In this case, the order will continue to exist in the LOB in some modified capacity. We now describe how the LOB handles this order modification. Without loss of generality, consider an ask $S_{i} \in \mathcal{S}$ with a quantity of $Q_{S_{i}}$ and a duration of $d_{S_{i}}$. With an incoming bid  $B_{j} \in \mathcal{B}$ with a quantity of $Q_{B_{j}} < Q_{S_{i}}$ and a duration of $d_{B_{j}} < d_{S_{i}}$ that matches with $S_{i}$ at time $t_{0}$, then $S_{i}$ will split into two different orders $S_{i1},S_{i2}$ with the \textbf{same price-time priority} as $S_{i}$. In particular, the LOB does the following:

    \begin{enumerate}
    
        \item $S_{i}$ becomes $S_{i1}$ with quantity $Q_{S_{i1}} = Q_{B_{j}}$ and $d_{S_{i1}} = d_{B_{j}}$ at time $t_{0}$ to fulfill the incoming buy order.
    
        \item An order for the residual from $S_{i}$ is placed as $S_{i2}$ with $Q_{S_{i2}} = Q_{S_{i}} - Q_{B_{j}}$ and $d_{S_{i2}} = d_{S_{i}} - Q_{B_{j}}$ at time $t_{0} + d_{B_{j}}$.

    \end{enumerate}

    While this modification scheme adds complexity to LOB operations (the LOB must maintain a queue to handle \textit{future} order submissions), it has several benefits. It ensures that even in the case of partial fulfillments, the full quantity and duration provided by a flexible order can be used. A system for future order placement also allows for the implementation of a potential futures/options market, although this extension is not considered further here. Finally, this scheme disincentivizes placing extremely short duration orders, leading to longer orders that provide stability in the LOB. If residual orders were not given the same price-time priority, then devices would not be incentivized to submit orders for large quantities for fear that any residuals would continuously get de-prioritized.
    %%% Just reference the example in the visualization section, need to show unmatched stacks, and matched stacks after stack cutting, and then show all the dispatches that will result

    % \subsection{PV Case Study}
    % Example of why PV always bids 0, splitting surplus, incentives created by doing that

    \begin{figure}[!t]

        \centerline{\includegraphics[width=0.95\columnwidth]{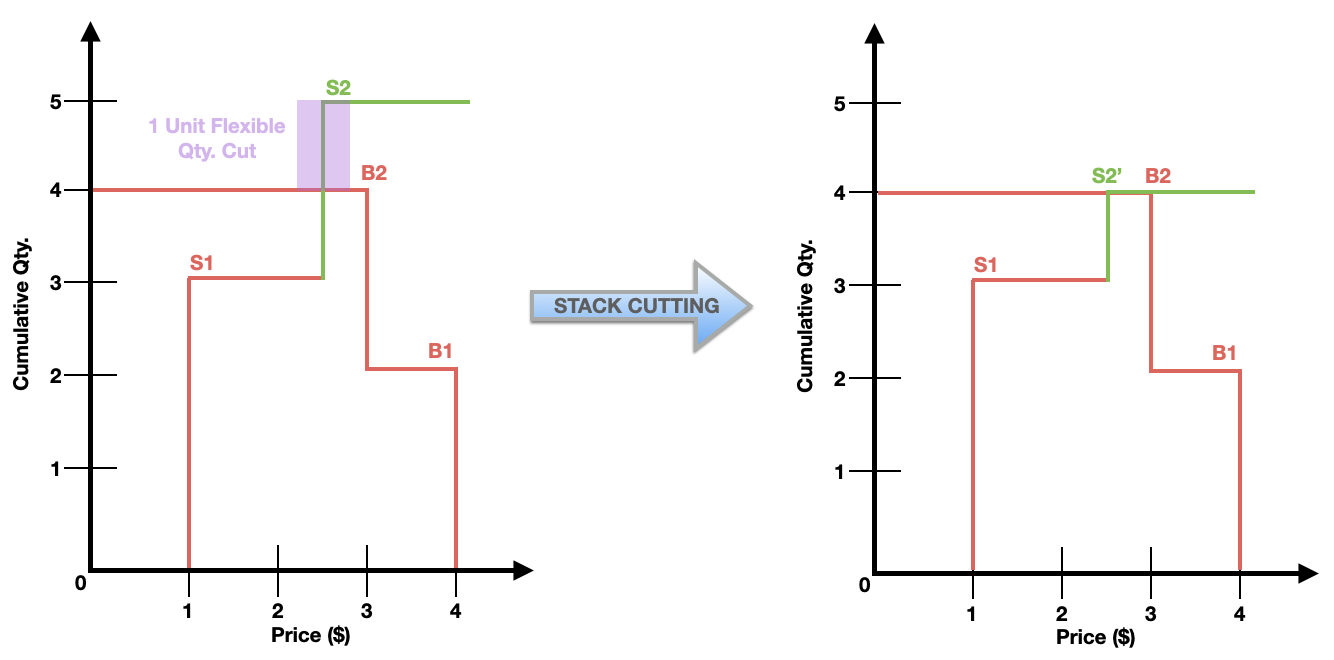}}

        \caption{1 unit of supply is cut from S2 (Order 3) to match the stacks. Flexible orders are green; inflexible orders are red.}
        \label{fig:StackCutting}

    \end{figure}
    
% \section{Orderbook Clearing and Settlement}
\section{Orderbook Settlement}
\label{sec:settlement}
    % \subsection{Clearing Mechanism}
    % The LOB tries to match the devices that bid negative quantity(sellers) with the devices that bid positive quantity(buyers) based on the limit price and priority of all the devices.
    % at given time t, let $f(X^{(t)}_{i}):= \{i, Q, P, F, T_{o}, T_{d}, T_{e}\}$ be a function that returns a vector of order details of an existing order from agent $i$ in the LOB. The orders are arranged in the LOB according to the price-time priority.
    
    % The orders are matched based on price and quantity and there are two cases in this scenario. 1. When the marginal unit is fully satisfies the order, then every order in the contract
    
    % If and suppose the LOB is changed from a new order or deletion of existing order. The matching engine tries to match orders based on the limit price, flexibility and position in the LOB. To visualize the clearing mechanism
    
    % Insert the graphical interface of clearing.
    
    \subsection{Settlement} 
    The \textit{ex post} cleared price that is discovered in each matching round is a single price determined by the marginal units. 
    
    \textbf{Case 1}: There is a single marginal unit cut. All devices in contract pay the marginal unit's price. 
    
    \textbf{Case 2}: The lowest buy order $B_{Low}$ and the highest sell order $S_{High}$ in contract are both marginal. All devices pay $\frac{B_{Low} + S_{High}}{2}$, thereby splitting the surplus. 
    
    Under this model, all the infra-marginal units receive a surplus and the price they pay is independent of their original limit price. The settlement model discourages devices from bidding strategically, rather encouraging them to submit an order that reflects their true cost of electricity. In a transactive system, all devices pay a tariff on top of the transmission cost for network usage.

%  Penality of devices non-conforming
    
    % The cleared devices pay the price of the marginal units.
    % Case 1: The price divides the bid quantity. Only one possible price.\\
    % Case 2: The quantity bid divides the price. Different price points are possible. We take the average price in the range.
    % In a contract where the agent's orders are matched and cleared, the lowest paying buy order and the highest paying sell order determines the settlement price for the other agents. The location where the marginal units intersects entails the price that every units pays and receives.
    
    % \subsection{Feeder Role} \label{sec:feeder}
     
    %  The feeder (utility) acts as a special type of market participant. Unlike other devices on the market, it does not employ strategic bidding to place LOB orders. Rather, it acts a pseudo-market maker providing liquidity in the market by always bidding the wholesale price of electricity in bid orders or the locational marginal price in ask orders. 
    % The utility does not do strategic bidding, but bids the wholesale price of electricity or the Locational Marginal Price. In case of the storage market, the utility does Market Order for the price of storage.

\section{Full Matching Algorithm Example}
\label{sec:example}

    We now apply the above algorithm on the LOB shown in Table~\ref{tab:LOBTableExample}. The spread is negative (-\$3.00). In this case, all the orders in $\mathcal{B}$ and $\mathcal{S}$ already match on price, so the result of the price-filtering yields $\mathcal{B'} = \mathcal{B}$ and  $\mathcal{S'} = \mathcal{S}$. We note that $S'_{TOT} \neq D'_{TOT}$, so we proceed with stack cutting.

    % \begin{figure}[!t]

    %     \centerline{\includegraphics[width=0.95\columnwidth]{LOB_Stacks.png}}

    %     \caption{The stacks ($\mathcal{B'}, \mathcal{S'}$) for the example in Table~\ref{tab:LOBTableExample}. $D'_{TOT} (10) \neq S'_{TOT} (12)$ so the 1 unit of supply cut is shown in gray.}
    %     \label{fig:StackExample}

    % \end{figure}

    Because there is a flexible order to cut from (Order~3), we are able to proceed as in the case of matched stacks after cutting 1 unit of quantity from Order~3. The stack cutting procedure is illustrated in Figure \ref{fig:StackCutting}. In this case, Device~3 is the marginal unit so all orders are dispatched at the marginal unit's price of \$2.50. The transactions of the form ($DeviceId_{seller}$, $DeviceId_{buyer}$, $Quantity$, $Price_{clearing}$) in the generated dispatch are listed in Table \ref{tab:Dispatch}.
    \begin{table}[t]
        \caption{Transactions from generated dispatch}
        \label{tab:Dispatch}
        \centering
        \begin{tabular}{|c|c|c|c|c|c|}
        \hline
        \textbf{Transaction} & \textbf{DeviceId} & \textbf{DeviceId} & \textbf{Quantity} & \textbf{Price (\$)} & \textbf{Duration} \\ \hline
        1 & 1 & 4 & 2 & 2.50 & 10 \\ \hline
        2 & 2 & 4 & 1 & 2.50 & 10 \\ \hline
        3 & 2 & 3 & 1 & 2.50 & 10 \\ \hline
        \end{tabular}

    \end{table}
    
    % DPC: suggest present this is a table so it's easier to read
    % \begin{enumerate}
    
    %     \item (1, 4, 6, \$12, 10)
    
    %     \item (2, 4, 2, \$12, 10)
    
    %     \item (2, 3, 4, \$12, 10)
    
    % \end{enumerate}

%% switch 

% TODO: Limit Orderbook mechanism alongside double auction is used in a Transactive Market and the efficiency of both should be compared. Lay the bases for economic experiments.Limit Orderbook exposes more in-depth demand elasticity which a traditional demand response program does not.

\section{Future Work}

     Key metrics that capture the performance of various market mechanisms need to be developed to quantify the advantages of the LOB over the time-delimited DA. These metrics fall into three categories: (i) quantifying the economic efficiency and surplus provided to all participants in a given market; (ii) characterizing control-theoretic response by modeling transactive energy markets as closed loop control systems; and (iii) understanding the performance of device bidding strategies that different market mechanisms enable.
    
    Additionally, several operational aspects are critical to implementing a LOB in a transactive energy system. For example, further study is warranted on ensuring robust guarantees between parties in contract to handle cancellations, trust, and imbalances. Potential solutions may include a penalty structure to disincentivize cancellations and promote trust between devices, as well as forcing devices to be fiscally responsible (via market orders) for procuring power. 
    % Further economic analysis is still required to quantify the economic efficiency and surplus provided to all market participants in a fully functioning LOB market. Furthermore, there is little work yet on the control-theoretic responsiveness of the LOB and modeling the orderbook as a closed loop control system. Finally, the individual bidding strategies of the devices that participate within the LOB are varied and complex and there remains significant work on how these devices can best interact with the market to achieve their goals. 
    
    % Further economic analysis is still required to quantify the economic efficiency and surplus provided to all market participants in a fully functioning LOB market. Furthermore, there is little work yet on the control-theoretic responsiveness of the LOB and modeling the orderbook as a closed loop control system. Finally, the individual bidding strategies of the devices that participate within the LOB are varied and complex and there remains significant work on how these devices can best interact with the market to achieve their goals. 

\section*{Conclusions}
\label{sec:conclusion}

    Transactive energy systems hold tremendous promise in coordinating DERs for more efficient distribution system operations. In this work, we have laid the framework for a limit orderbook market mechanism that can compensate for the limitations of the time-delimited double auction typically used in transactive energy. We have provided algorithms and rules for market operation that ensure efficient and robust markets.

\section*{Acknowledgment}

    This work was funded by the US Department of Energy's Office of Energy Efficiency and Renewable Energy Connected Communities program under Contract DE-AC02-76SF00515.

% \section*{References}

    % Please number citations consecutively within brackets \cite{b1}. The 
    % sentence punctuation follows the bracket \cite{b2}. Refer simply to the reference 
    % number, as in \cite{b3}---do not use ``Ref. \cite{b3}'' or ``reference \cite{b3}'' except at 
    % the beginning of a sentence: ``Reference \cite{b3} was the first $\ldots$''
    
    % Number footnotes separately in superscripts. Place the actual footnote at 
    % the bottom of the column in which it was cited. Do not put footnotes in the 
    % abstract or reference list. Use letters for table footnotes.
    
    % Unless there are six authors or more give all authors' names; do not use 
    % ``et al.''. Papers that have not been published, even if they have been 
    % submitted for publication, should be cited as ``unpublished'' \cite{b4}. Papers 
    % that have been accepted for publication should be cited as ``in press'' \cite{b5}. 
    % Capitalize only the first word in a paper title, except for proper nouns and 
    % element symbols.
    
    % For papers published in translation journals, please give the English 
    % citation first, followed by the original foreign-language citation \cite{b6}.

\end{document}